\documentclass[11pt]{article}
\usepackage[utf8]{inputenc}

\usepackage{jheppub}

\usepackage{amsfonts}
\usepackage{amsthm}
\usepackage{mathtools}
\usepackage{slashed}
\usepackage{tensor}
\usepackage{physics}

\usepackage{float}
\usepackage{array}

\usepackage{color}
\usepackage{enumitem}
\usepackage{lipsum}

\definecolor{Dgreen}{RGB}{45,130,85}

\newcommand{\nn}{\nonumber}
\newcommand{\secterm}[1]{\qty{#1}_{\text{sec.}}}

\newcommand{\RR}{\mathbb{R}}
\newcommand{\ZZ}{\mathbb{Z}}
\newcommand{\NN}{\mathbb{N}}

\newcommand{\cD}{\mathcal{D}}
\newcommand{\cF}{\mathcal{F}}

\newcommand{\cL}{\mathcal{L}}

\newcommand{\cN}{\mathcal{N}}
\newcommand{\cO}{\mathcal{O}}

\newcommand{\cX}{\mathcal{X}}

\newcommand{\ads}{\ensuremath{\mathrm{AdS}}}
\newcommand{\adst}{\ensuremath{\mathrm{AdS}_3}}

\begin{document}
	\title{Time-dependent microstrata in \boldmath\adst}

	\author[a]{Anthony Houppe}
	\affiliation[a]{Institut für Theoretische Physik, ETH Zürich, \\ Wolfgang-Pauli-Strasse 27, 8093 Zürich, Switzerland}
	\emailAdd{ahouppe@phys.ethz.ch}

	\abstract{We use perturbation theory to construct a family of time-dependent microstrata: a set of non-extremal solutions of IIB supergravity asymptotic to $\mathrm{AdS}_3 \times S^3 \times T^4$. Our construction shows that the ``special locus'' constraints of \cite{ganchev.giusto.ea:2023} can be broken by allowing the solutions to depend on time. We study the secular terms appearing in the perturbation theory. Some of them can be resummed into frequency shifts, with the same interpretation as for the previously-studied microstrata solutions. Other secular terms appear harder to resum, questioning the long-term stability of the solutions.}

	\maketitle
	\flushbottom

	\section{Introduction}
	\label{sec:Intro}

	Over the years, many supersymmetric microstate geometries of the three-charge black hole have been constructed (see \cite{bena.martinec.ea:2022,bena.martinec.ea:2022*1} for reviews). These geometries can often be matched explicitly with the BPS states of the underlying D1-D5 brane system. Among those, superstrata constitute a large family \cite{bena.giusto.ea:2015,bena.martinec.ea:2016,bena.giusto.ea:2016,bena.giusto.ea:2018,ceplak.russo.ea:2019,heidmann.warner:2019,ganchev.houppe.ea:2022*1,heidmann.mayerson.ea:2020,ceplak:2022}, whose CFT dual are heavy states of the D1-D5 CFT with momentum excitations \cite{giusto.mathur.ea:2004,giusto.mathur.ea:2005,giusto.lunin.ea:2013,rawash.turton:2021}.

	One of the challenges of the microstate geometry programme is to go beyond supersymmetry: to address the information paradox, one needs to build geometries that can decay and emit Hawking radiation. Here the challenge is two-fold. From the bulk perspective, one has to solve the Einstein equations at strong coupling, without the crutch provided by the BPS equations. From the point of view of holography, the renormalization group flow between the supergravity regime and the zero-coupling regime can be very complicated and makes the matching with the CFT states a much harder task.

	Nonetheless, several families of non-supersymmetric geometries have been constructed recently. Bubbled geometries \cite{Bah:2023ows,heidmann:2022,bah.heidmann.ea:2022} are explicit solutions far from the BPS regime, but whose CFT duals are not yet identified. The focus of this paper is on microstrata \cite{ganchev.houppe.ea:2021,ganchev.giusto.ea:2022,ganchev.giusto.ea:2023}: they are non-BPS generalizations of superstrata, and as such their CFT duals have been identified. While there is no explicit formulation of these geometries far from the BPS regimes, they have been studied both perturbatively and numerically to a high precision.

	Superstrata and microstrata are both solutions of Type IIB supergravity, they are the backreaction of a system of D1 and D5 branes, sharing a common direction, with momentum excitations. The vacuum of this system is global $\adst \times S^3 \times T^4$. From there, one can add left-moving excitations that preserves some amount of supersymmetry to obtain superstrata. If one adds both left-moving and right-moving excitations, the solution will generically break all supersymmetries: these are microstrata solutions.

	Trying to solve directly the 10-dimensional equations of motion without supersymmetry is an impossible task ; one needs to simplify the problem first. Even when working with supersymmetry, one mostly considers solutions that do not have any excitations in the $T^4$ directions, and thus can be studied in a truncated six-dimensional supergravity theory: this is the theory where superstrata live \cite{bena.giusto.ea:2015}. But one can go a step further, and restrict ones attention to a small subset of the degrees of freedom on the $S^3$ as well. Such a truncation results in a theory of gauged supergravity in three dimensions, with gauge group $SO(4)$. This theory can describe a subfamily of the superstrata solutions, namely the $(1,m,n)$-superstrata, and has been instrumental in the construction of microstrata \cite{ganchev.houppe.ea:2021}. It has been proved to be consistent: its solutions can be uplifted to \emph{exact} solutions of Type IIB supergravity \cite{mayerson.walker.ea:2020}.

	The previous studies of microstrata focused on solutions belonging to a ``Q-ball'' ansatz that are, in a way, ``time independent''. The meaning of time independence is a bit subtle, because it is not a gauge-invariant concept. In this context, it means that there exists a gauge in which the fields do not have any explicit dependence in time. Because the Cartan subalgebra of the gauge group $SO(4)$ is $u(1) \oplus u(1)$, such solutions can only depend on time through two independent ``frequencies''\footnote{While these frequencies appear classically to be gauge artifacts, one must remember that the large gauge transformations are anomalous in the quantum theory, and so these frequencies acquire definite values and a physical meaning.}. The other frequencies are then locked in a two-to-one ratio with the independent frequencies.

	This phase locking mechanism has been found in \cite{ganchev.giusto.ea:2022,ganchev.giusto.ea:2023} to severely restrict the explorable phase space of microstrata. Microstrata are built as the backreaction of some scalar excitations, and while their amplitudes should a priori be independent, regular solutions have only been obtained when these amplitudes respect quadratic constraints. These constraints have been dubbed ``special locus''. In the dual CFT, these particular points in the moduli space are identified with multi-particle states that are constructed via OPEs of a large number of single-particle states. 

	In this paper, we endeavor to construct microstrata away from the special locus, to explore a greater part of the phase space of microstrata. It was hypothesized in \cite{ganchev.giusto.ea:2023} that doing so requires to break time independence. After devising a new, more general ansatz for the fields that allow an explicit dependence on time, we will use perturbation theory to construct these solutions.

	Breaking time independence means that one is free to include many more scalar excitations. This paper will focus on the same excitations as were previously studied in \cite{ganchev.giusto.ea:2023}, to demonstrate that one can break away from the special locus. In particular we will see that, starting with time-independent excitations, the fields acquire an explicit time-dependence at third order in perturbation theory.

	A common pitfall when dealing with time-dependent perturbation theory in \ads{} is the presence of \emph{secular terms} \cite{chen:1995ena,bizon_weakly_2011,balasubramanian_holographic_2014,craps_renormalization_2014}. These terms grow unboundedly with time, and at large enough times break the perturbation theory. We will explore ways to resum these terms. As we will see, some of them can be reabsorbed by frequency shifts, similar to what was found for the time-independent solutions. Contrary to the latter, these shifts make the frequencies break the phase locking imposed by time-independence, explaining why they could not be constructed within the Q-ball ansatz. The non-linear interactions  will mix these shifted frequencies at higher orders, and because these shifts also represent shifts in energies, the set of fully backreacted solutions is expected to have a chaotic spectrum.

	In Section~\ref{sec:3dsuper}, we begin with a short review of the three-dimensional supergravity theory that is used to build microstrata. We then design an ansatz for the fields of the theory, inspired from \cite{ganchev.houppe.ea:2021,ganchev.giusto.ea:2023}, but including explicit time-dependence. We obtain a system of 9 real and 2 complex functions of two variables: time and the radial coordinate, governed by a set of partial differential equations. We also discuss the conditions on these functions to obtain a regular solution.

	In Section~\ref{sec:naive_pert} we construct microstrata using a naive perturbation theory. We start from the same perturbations as in \cite{ganchev.giusto.ea:2023}. Instead of imposing the special locus constraints, we find that one can build regular solutions by introducing a time-dependence at third order in perturbation theory. These additional terms are however secular, and break the regime of validity of perturbation theory to small time scales. We explain how to resum these terms to shifts in the frequencies.

	In Section~\ref{sec:poincare_method}, we explore a more systematic way of resumming secular terms, using the Poincaré-Lindstedt method. We find that this method fails at fourth order in perturbation theory, and discuss the possible causes of this failure.

	Finally, in Section~\ref{sec:final_comments} we comment on the results and explore further directions. The paper also contains two appendices, \ref{sec:second_order_perturbation_theory} and \ref{sec:second_order_poincare_lindstedt}, detailing some computations of the perturbation theories.

	\section{Review of the three-dimensional supergravity}
	\label{sec:3dsuper}

	In this section, we summarize the three dimensional supergravity that was developed in \cite{mayerson.walker.ea:2020}, and used for the construction of microstrata \cite{houppe.warner:2020,ganchev.houppe.ea:2021,ganchev.giusto.ea:2022,ganchev.giusto.ea:2023}. We then describe a new ansatz for the fields, somewhat different from the one used in these previous works. This ansatz will be the basis for the construction of a new class of microstrata that explicitly depends on time.

	\subsection{The bosonic action} 
	\label{sub:the_bosonic_field_content}
	
	The theory of interest is $\cN = 4$ gauged supergravity in three dimensions, with gauge group $SO(4)$. It was first studied in \cite{mayerson.walker.ea:2020}, where it was shown to be a consistent truncation of a six-dimensional $\cN=2$ theory, that is useful to describe superstrata. The theory can be understood as a particular near-brane limit of a system of coincident D1 and D5 branes in Type IIB supergravity. The aim of this section is not to completely characterize this theory, as this has been done on several occasions \cite{mayerson.walker.ea:2020,houppe.warner:2020,ganchev.houppe.ea:2021}. Instead, this is a brief summary of everything that is required to construct microstrata. In particular we will not explore the fermionic content of the theory.

	The theory has a gauge group $SO(4)$, and we will use capital latin indices, $I, J$, to denote the vector representation of $SO(4)$. The fields comprise:
	\begin{itemize}
		\item A three-dimensional metric $g_{\mu\nu}$, with signature $(+ - -)$, following the conventions established in \cite{houppe.warner:2020}.
		\item A vector of scalars $\chi_I$.
		\item A symmetric matrix of scalars $m_{IJ}$.
		\item An antisymmetric matrix of gauge fields $A_\mu^{IJ}$, and its dual $\tilde A_\mu^{IJ} \equiv \frac12 \epsilon_{IJKL} A_\mu^{KL}$, with minimal couplings  $$\cD_\mu \, \cX_{I}  ~=~ \partial_\mu \, \cX_{I}   ~-~  2\, g_0\,\widetilde A_\mu{}^{IJ} \, \cX_{J} \,.$$
	\end{itemize}

	The gauge coupling is denoted by $g_0$, and is related to the charges of D1-D5 system:
	$g_0 = (Q_1 Q_5)^{-1/4}$.
	We will, however, eventually rescale the fields so that this coupling is factored out of the equations of motion.

	The (bosonic part of the) action of the theory can be written as
	\begin{equation}
			\begin{aligned}
				\cL ~=~ & -\frac{1}{4} \,e\,R    ~+~ \frac{1}{8}\,e \, g^{\mu \nu} \, m^{IJ}   \, (\cD_\mu\, \chi_{I})  \, (\cD_\nu\, \chi_{J})  ~+~  \frac{1}{16}\,e \, g^{\mu \nu} \,  \big( m^{IK} \, \cD_\mu\, m_{KJ}  \big)   \big( m^{JL} \, \cD_\nu\, m_{LI}  \big)  \\
				&+~  \frac{1}{2}\,e  \, \varepsilon^{\mu \nu \rho} \, \Big[  g_0 \,\big(A_\mu{}^{IJ}\, \partial_\nu  \widetilde A_\rho{}^{IJ}  ~+~\frac{4}{3}\,  g_0 \, A_\mu{}^{IJ} \,  A_\nu{}^{JK}\, A_\rho{}^{KI} \,\big) ~+~  \frac{1}{8}\,  {Y_\mu}{}^{IJ}  \, F_{\nu \rho}^{IJ}  \\
				&-~   \frac{1}{8}\, e \, g^{\mu \rho}  \, g^{\nu \sigma} \, m_{IK} \,m_{JL}\,  F_{\mu \nu }^{IJ}  \, F_{\rho \sigma }^{KL}  ~-~e\, V
				 \Big] \,,
			\end{aligned}
			\label{eq:3Daction}
		\end{equation}
	where $e = \sqrt{\det g}$ and we have introduced the scalar current
	\begin{equation}
	    Y_{\mu \, IJ}  ~\equiv~  \chi_J \,\cD_\mu \chi_I ~-~  \chi_I \,\cD_\mu\chi_J  \,,
	 \end{equation}
	and the scalar potential
	\begin{equation}
		V ~=~  \frac{1}{4}\, g_0^2   \,  \det\big(m^{IJ}\big) \, \Big [\, 2 \,\big(1- \frac{1}{4} \,  (\chi_I \chi_I)\big)^2    ~+~ m_{IJ} m_{IJ}  ~+~\frac{1}{2} \,  m_{IJ} \chi_I \chi_J  ~-~\frac{1}{2} \,  m_{II}  \,  m_{JJ}\, \Big] \,.
		\label{eq:potential1}
	\end{equation}

	The supersymmetric vacuum of the theory is global \adst, with all the scalar and gauge fields set to zero, and the \ads{} radius is given by the coupling $R_{\ads} = g_0^{-1}$.

	\subsection{A time-dependent ansatz} 
	\label{sub:a_time_dependent_ansatz}
	
	In this section, we devise a tractable ansatz for the fields, following the procedure laid out in \cite{ganchev.houppe.ea:2021,ganchev.giusto.ea:2023}. We will mostly use the same conventions as in these papers.\footnote{Departures from said conventions will be stated explicitly.} The key difference with respect to \cite{ganchev.giusto.ea:2023} is that we will allow the fields to depend explicitly on time. We will also restrict the theory to a specific $U(1)$-invariant sector, similarly to the choice made in \cite{ganchev.houppe.ea:2021}.

	Concretely, we impose the following conditions to the fields:
	\begin{enumerate}[label=(\roman*)]
		\item Invariance under translations of the \ads-circle direction ;
		\item Invariance under the internal $U(1) \subset SO(4)$ that rotates the indices $(3,4)$ ;
		\item Reflection invariance under $(\sigma \to - \sigma, \, t \to -t)$, where $t$ is time and $\sigma$ is the \ads-circle coordinate, along with a discrete $SO(4)$ rotation $(2 \to -2, \, 4 \to -4)$.
	\end{enumerate}

	It is easy to check that these choices are compatible with the equations of motion, and thus lead to a consistent truncation. We can now examine the fields one by one and devise an ansatz for them that respects these constraints.

	\subsubsection{The metric} 
	\label{sub:the_metric}

	We are interested in solutions to the equations of motion that are smooth, local deformations of the \adst{} metric. This condition of smoothness in three dimensions will be used to discrimate physical solutions. Note that technically, one should look at the smoothness of the solution once uplifted in six dimensions. However, we expect that singular solutions in 3D are generically singular in 6D as well: the contrary would require precise cancellations between various fields (as can be seen using the uplift formulas of \cite{mayerson.walker.ea:2020}).\footnote{In particular, the log-singularities that are the focus of the next sections are not cured after the uplift, and thus solutions that include those singularities are unphysical.}

	Recall that the metric of global \adst{}  with radius $R_{\ads}$ can be written as
	\begin{equation}
		ds_{\adst}^{2}  ~=~ R_{\ads}^2 \, \bigg[ \,   \big(\rho^2 +1\big)\,  dt ^2~-~  \frac{d\rho^2}{\rho^2 + 1}  ~-~ \rho^2\,  d\sigma^2 \, \bigg] \,,
		\label{eq:AdSmet}
	\end{equation}
	where the circle coordinate $\sigma$ has periodicity $2\pi$. It will prove useful to make the change of coordinates:
	\begin{equation}
		\xi ~=~\frac{\rho}{\sqrt{\rho^2+1}} \,,  \qquad  \psi~=~ t+\sigma\,, 
		\label{xidef}
	\end{equation}
	so that the new radial coordinate $\xi$ is compact, $0 < \xi < 1$, and $\psi$ is a light-cone coordinate. In these coordinates, the metric of global \ads{} becomes
	\begin{equation}
		ds_{\adst}^{2}  ~=~  R_{\ads}^2 \, \bigg[  \, \bigg(d t +   \frac{\xi^2}{(1- \xi^{2})} \, d\psi \bigg)^2~-~\,\frac{1}{(1-\xi^{2} )^{2}} \, \big( d \xi^2 ~+~ \xi^2 \, d \psi^2 \big) \, \bigg] \,.
	\end{equation}

	The ansatz that we choose for the metric is
	\begin{equation}
		ds_{3}^{2}  ~=~  R_{\ads}^2 \, \bigg[ \,\Omega_1^{2} \, \bigg(d t +   \frac{k}{(1- \xi^{2})} \, d\psi \bigg)^2~-~\,\frac{\Omega_0^{2}}{(1-\xi^{2} )^{2}} \, \big( d \xi^2 ~+~ \xi^2 \, d \psi^2 \big) \, \bigg] \,,
		\label{eq:metric1}
	\end{equation}
	for three arbitrary (smooth) functions $\Omega_0$, $\Omega_1$ and $k$, that depend a priori on all three coordinates $(t, \psi, \xi)$. Setting $\Omega_0 = \Omega_1 = 1$, and $k = \xi^2$, one recovers the metric of global \ads. 

	These three functions are subject to several constraints, stemming from the requirements of smoothness, fixed \ads{} asymptotics, and from the condition (i). First, (i) imposes that the three fields do not depend on $\psi$. At the origin, $\xi = 0$, the metric is regular if 
	 \begin{equation}
	     k = \cO(\xi^2) \qq{as} \xi \to 0 \,. 
	     \label{eq:kat0}
	 \end{equation}
	We must also require that the $\psi$-circle is everywhere space-like, and this leads to the condition
	\begin{equation}
	    \xi^2 \Omega_0^2 - \Omega_1^2 k^2 \geq 0 \,.
	     \label{eq:noCTC}
	\end{equation}
	The last constraint come from the fact that the leading divergences in $d\psi^2$ at infinity are of order $(1-\xi^2)^{-2} \sim \rho^4$. For the metric to be asymptotic to \adst, they must cancel, which implies that
	\begin{equation}
	    \Omega_1(\xi = 1) = \frac{\Omega_0(\xi = 1)}{k(\xi = 1)} \,.
	     \label{eq:omegaat1}
	\end{equation}
	Now that all the constraints are satisfied, there remains to fix a gauge invariance. The form of the metric $\eqref{eq:metric1}$ contains an unfixed rescaling of the time coordinate, $t \to c \,t$, which has the effect of rescaling the functions as $\Omega_1 \to c \,\Omega_1, \, k \to c k$. Following \cite{ganchev.giusto.ea:2023}, we choose to fix this rescaling by imposing
	\begin{equation}
	    \frac{g_{\sigma\sigma}}{g_{tt}} \xrightarrow[\xi \to 1]{} -1 \,,
	     \label{eq:scaleat1}
	\end{equation}
	which is equivalent to asking that the asymptotic metric is sectioned by cylindrical Minkowski slices, $ds_2^2 \sim \rho^2 (dt^2 - d\sigma^2)$.

	\subsubsection{The scalars} 
	\label{ssub:the_scalars}
	
	First, because of the condition (i), all the fields are functions of $t$ and $\xi$ only.
	For the vector field $\chi_I$, the symmetry (ii) implies that $\chi_3 = \chi_4 = 0$. The symmetry (iii) implies that $\chi_1$ is an even function of time, $\chi_1(-t) = \chi_1(t)$, while $\chi_2$ is an odd function, $\chi_2(-t) = -\chi_2(t)$. We will parametrize them in the following way:\footnote{We introduce the factor $\sqrt{1-\xi^2}$ because, as we will see, the solutions to the linearized equation of motion for the functions $\nu_1$ and $\nu_2$ are then polynomials in $\xi$. Note also that the field that we name $\nu_2$ is different from the similarly named field in \cite{ganchev.giusto.ea:2023}.
}
	\begin{equation}
	    \chi_1 = \sqrt{1-\xi^2} \,\nu_1(t, \xi) \,,\quad \chi_2 = \sqrt{1-\xi^2} \,\nu_2(t, \xi) \,.
	\end{equation}
	We also introduce a complex field
	\begin{equation}
		\nu(t, \xi) \equiv \nu_1(t, \xi) + i \,\nu_2(t, \xi) 
	\end{equation}
	and the symmetry (iii) then implies that $\nu(-t, \xi) = \overline{\nu(t, \xi)}$.
	
	We now turn to the matrix $m$. Because of (ii), the only non-zero terms are $m_{11},m_{22},m_{12}=m_{21}$ and $m_{33}=m_{44}$. We parametrize it as:
	 \begin{equation}
	     m = \mqty( e^{\mu_1} + \lambda_1 & \lambda_2 & 0 & 0 \\
	     			\lambda_2 & e^{\mu_1} - \lambda_1 & 0 & 0 \\
	     			0 & 0 & e^{\mu_2} & 0 \\
	     			0 & 0 & 0 & e^{\mu_2}) \,,
	 \end{equation}
	 where $\mu_i, \lambda_i, i = 1,2$ are functions of $t$ and $\xi$. Note that the definition of $\lambda_2$ is different from the one in \cite{ganchev.giusto.ea:2023}. The symmetry (iii) imposes that $\lambda_2$ is an odd function of $t$, while the others are even. We will use the notation $\mu_{\pm} \equiv \mu_1 \pm \mu_2$, and we introduce again a complex field 
	 \begin{equation}
	     \lambda(t, \xi) ~\equiv~ m_{11} - m_{22} + i \, m_{12} ~=~ \lambda_1(t, \xi) \,+\, i \,\lambda_2(t, \xi) \,,
	 \end{equation}
	and the symmetry (iii) implies that $\lambda(-t, \xi) = \overline{\lambda(t, \xi)}$.

	We also impose that the scalars $\chi_I$ and $m_{IJ}$ vanish at infinity, which translates to
	\begin{equation}
	    \nu(\xi = 1) \sim \cO(1) \,, \quad \lambda(\xi = 1) = 0 \,, \quad  \mu_{1,2}(\xi = 1) = 0 \,.
	    \label{eq:scalarsat1}
	\end{equation}

	\subsubsection{The gauge fields} 
	\label{ssub:the_gauge_fields}
	

	Only two components of the gauge fields survive the condition (ii), $\tilde A^{12}$ and $\tilde A^{34}$, reducing the gauge group to $U(1) \times U(1)$. By a gauge transformation, we can suppress their $d\xi$ components, thus we can write them as:
	\begin{equation}
		\tilde A^{12} ~=~ \frac{1}{g_0} \,\big[\,  \Phi_1(t, \xi)  \, dt ~+~  \Psi_1(t, \xi)  \, d\psi \, \big]\,, \qquad  \tilde A^{34} ~=~ \frac{1}{g_0} \,\big[\,\Phi_2(t, \xi)  \, dt ~+~  \Psi_2(t, \xi)  \, d\psi    \, \big] \,,
	\end{equation}
	where $\Phi_i, \Psi_i, i=1,2$ are functions of $t$ and $\xi$, and are even functions of time because of (iii). We have introduced the factors of $g_0$ to make the equations of motion scale independent.

	These functions must be smooth over the whole spacetime, and the periodicity of the $\psi$-circle further requires to fix the magnetic gauge fields at the origin. It would be natural to impose that they vanish at the origin, but we will actually need a more general condition:
	\begin{equation}
	    \Psi_1(\xi=0) = \frac{n}{2} \,, \quad \Psi_2(\xi = 0) = 0 \,,
	    \label{eq:magneticat0}
	\end{equation}
	where $n \in \ZZ$. It may appear strange, as solutions with $n \neq 0$ are not strictly-speaking smooth. We still consider them, because they are dual through a $\psi$-dependent $U(1)$ gauge transformation to solutions that are smooth (see \cite{ganchev.giusto.ea:2023} for more details). In other words, the smooth solutions that we are looking for do not fit directly the ansatz because of their $\psi$-dependence, but their gauge-rotated equivalent do. Once those are obtained, undoing the rotation is trivial.

	We understand the integer $n$ as the magnetic \emph{mode number} of the solutions. The ansatz only contains single-mode solutions: the $\psi$-dependence of a multi-mode solution cannot be gauged away.

	There remains to fix two $U(1)$ global gauge choices. With the fields of this ansatz, the $U(1)_1$ gauge transformation is:
	\begin{equation}
		\nu \to e^{i \delta t} \nu \,,\quad \lambda \to e^{2i \delta t} \lambda \,,\quad \Phi_1 \to \Phi_1 - \delta/2 \,,
		\label{eq:u1gauge}
	\end{equation}
	for any $\delta \in \RR$. Note the fact that $\lambda$ rotates with twice the frequency of $\nu$. The second symmetry $U(1)_2$ simply only shifts $\Phi_2$. We can fix those two gauges by imposing:
	\begin{equation}
	    \Phi_1(t=0,\xi=0) = 0 \,,\quad \Phi_2(t=0,\xi=0) = 0 \,.
	    \label{eq:electricat0}
	\end{equation}

	\vspace{.5em}
	To summarize, the full ansatz involves 2 complex and 9 real functions of two variables, the time $t$ and the radius $\xi$:
	\begin{equation}
	    \cF = \{ \nu,\lambda,\mu_1,\mu_2,\Phi_1,\Phi_2,\Psi_1,\Psi_2,\Omega_0,\Omega_1,k \} \,,
	    \label{eq:list_func_ansatz}
	\end{equation}
	and these functions are subject to the conditions \eqref{eq:kat0}, \eqref{eq:noCTC}, \eqref{eq:omegaat1}, \eqref{eq:scaleat1}, \eqref{eq:scalarsat1}, \eqref{eq:magneticat0} and \eqref{eq:electricat0}.

	Note that, up to the choice of gauge \eqref{eq:electricat0}, this ansatz is a strict superset of the one used in \cite{ganchev.houppe.ea:2021}. The solutions of \cite{ganchev.houppe.ea:2021} can be recovered by fixing
	\begin{equation}
		\nu = \tilde\nu(\xi) e^{i \omega t}\,,\quad
		\lambda = e^{2\tilde\mu_1(\xi)}\sinh(2\tilde\mu_0(\xi)) e^{2i\omega t} \qand 
		e^{\mu_1} = e^{2\tilde\mu_1(\xi)}\cosh(2\tilde\mu_0(\xi))\,,
	\end{equation}
	 for some constant $\omega$ ($=2$ in \cite{ganchev.houppe.ea:2021}), by fixing $\Phi_1(\xi = 0) = 1$, and by removing the time-dependence of all the other fields. The ansatz of \cite{ganchev.giusto.ea:2023}, however, contains fields that are not present here.

	\section{Naive perturbation theory}
	\label{sec:naive_pert}

	While microstrata cannot be computed analytically, they can be studied using perturbation theory. We focus on the backreaction of very particular excitations: they are excitations of the complex scalars $\nu$ and $\lambda$, and are respectively referred to as $\alpha$- and $\beta$-class.

	Following the procedure of \cite{ganchev.giusto.ea:2023}, we start by fixing  the background to be  empty global \adst, which translates in terms of the fields of the ansatz to:
	\begin{equation} 
	\begin{gathered}
		\nu^{\ads} = 0 \,,\qquad \lambda^{\ads} = 0 \,,\qquad \Omega_0^{\ads} = \Omega_1^{\ads} = 1 \,,\qquad k^{\ads} = \xi^2 \,,\\
		\Psi_1^\ads = \frac{n}2 \,,\qquad \Psi_2^\ads = 0 \,,\qquad \Phi_1^\ads = \Phi_2^\ads = 0 \,,
	\end{gathered}
	\label{eq:ads_background_fields}
	\end{equation}
	for some fixed magnetic mode number $n \in \ZZ$. We then expand all the fields as
	\begin{equation}
	    X(t,\xi) = X^\ads + \sum_{k=1}^\infty \epsilon^k \, \delta^{(k)}\! X (t,\xi) \,, \quad X \in \cF \,,
	\end{equation}
	where $\epsilon$ acts as a bookkeeping parameter for the perturbation theory. By substituting this expansion in the equations of motion and equating the powers of $\epsilon$, we obtain the perturbative equations at each order.

	We start at order $\epsilon^1$, by choosing a normalizable excitation. The equations of motion then completely fix the higher orders, in terms of this choice and of the integer $n$. Keeping $n$ as a free parameter, we will compute, in Section~\ref{sec:time_dep_solutions}, the backreaction of a two-mode excitation up to third order. For specific choices of $n$, one can compute the solutions to much higher orders in a systematic way: as an example and further check of the validity of the perturbative expansion, we have computed a time-dependent solution at $n = 2$ up to 10th order in $\epsilon$. 

	\subsection{The linearized equations} 
	\label{sub:linearized_equations}

	At order $\epsilon^1$, the linearized equations of motion of the two complex scalars in the \adst{} background are
	\begin{align}
	    \xi^2 \qty(\partial_t^2 \nu + 2 i n \,\partial_t \nu) ~&=~ \xi \partial_\xi \qty(\xi \qty(1-\xi^2) \partial_\xi \nu) \,-\, \qty(\xi^2 + n^2 \qty(1 - \xi^2)) \nu \,,
	    \\
	    \xi^2 \qty(\partial_t^2 \lambda + 4 i n \,\partial_t \lambda) ~&=~ (1-\xi^2) \, \xi \partial_\xi \qty(\xi\partial_\xi \lambda) \,+\, 4 \, n^2 (1-\xi^2) \lambda \,.
	\end{align}
	The solutions to these equations are (linear combinations of) waves of the form:
	\begin{equation}
	\begin{aligned}
		\nu ~&=~ \alpha\, \xi^{n}\, {}_2 F_1\qty(\frac{1-\omega}{2},\, n + \frac{1+\omega}{2},\, 1+ n \,;\, \xi^2) e^{i \omega t} \,,
		\\
		\lambda ~&=~ \beta\, \xi^{2 n} \, {}_2 F_1\qty(-\hat\omega,\, 2 n + \hat\omega,\, 1+ 2n \,;\, \xi^2) e^{2 i \hat\omega t} \,,
	\end{aligned}
	\label{eq:wave_solutions}
	\end{equation}
	where $\alpha$ and $\beta$ denote the amplitudes of the given modes. These solutions are regular and normalizable provided that the frequencies are chosen so that $\omega \in 2 \NN + 1$ and $\hat\omega \in \NN$.

	\subsection{The special locus}
	\label{sub:sploc}

	A subset of the excitations \eqref{eq:wave_solutions} has been studied in \cite{ganchev.houppe.ea:2021,ganchev.giusto.ea:2023}, by using restrictive, time-independent, ansätze. In our context, time-independence is understood to mean that there exist a gauge in which no field has an explicit dependence in time. The excitations \eqref{eq:wave_solutions}, in particular, are time-independent if $\omega = \hat\omega$. Indeed, under a $U(1)$ gauge rotation, the scalar $\lambda$ rotates with twice the frequency as the scalar $\nu$, and this is reflected by the factor $2$ in the phase of $\lambda$.

	While these excitations are well inscribed within the time-independent ansätze, it was found in \cite{ganchev.houppe.ea:2021,ganchev.giusto.ea:2023} that generically, their backreactions are not.\footnote{One exception is the case $\omega = \hat\omega = 1$. These solutions are supersymmetric, and their backreaction is computed analytically in \cite{ganchev.houppe.ea:2022*1}.} Nonetheless, two families of solutions have been identified. The first, so-called ``$\beta$-class solutions'', correspond to the backreaction of a single mode of the $\lambda$ scalar. They are obtained by setting $\alpha = 0$ and keeping $\beta$ arbitrary. The scalar $\nu$ vanishes in the full backreacted solutions.

	The second family of solutions is obtained by studying the backreaction of a single mode in $\nu$. Interestingly, the scalar $\lambda$ does not vanish. In fact, it is excited with an amplitude that depends quadratically on the amplitude of the excitation of $\nu$:
	\begin{equation}
		\beta = - \frac{\alpha^2}4 \,.
	\end{equation}
	This relationship between the amplitudes of the $\nu$ and $\lambda$ modes has been first discovered numerically in \cite{ganchev.houppe.ea:2021}, and has been dubbed the \emph{special locus}. Its existence has been confirmed using perturbation theory in \cite{ganchev.giusto.ea:2022}, and it has then been thoroughly examined in \cite{ganchev.giusto.ea:2023}.

	A major question is to understand the fate of the excitations whose amplitudes are not on the special locus. Trying to construct them using time-independent perturbation theory leads to incurable log-singularities, a sign that the solutions are not physical. One can conjecture a reason for this phenomenon by studying an  interesting property of the special locus solutions. For these, it was noted that the frequency, $\omega=\hat\omega$, of the scalars $\nu$ and $\lambda$ shifts non-linearly with the amplitude $\alpha$ of the perturbation. Such frequency shifts are a staple of the breaking of supersymmetry, and are expected to be present in all backreacted solutions.

	However, for a generic solution, there is a priori no reason to expect that these shifts preserve the relation $\omega = \hat\omega$. This equality is necessary for the solution to be time-independent, as it ensures the two-to-one ratio between the frequencies of $\nu$ and $\lambda$. We are lead to the conclusion that even though the excitations are time-independent, the backreacted solutions are generically time-dependent.

	In the next section, we confirm this intuition by building these backreacted solutions in the time-dependent ansatz of section~\ref{sec:3dsuper}.

	\section{Time-dependent solutions}
	\label{sec:time_dep_solutions}

	The objective is to compute the backreaction of the excitations \eqref{eq:wave_solutions} within perturbation theory. As described in section~\ref{sub:sploc}, we want to understand the backreaction of the time-independent excitations, thus we fix $\omega = \hat\omega$. We will keep the parameters $\alpha$ and $\beta$ generic and non-zero, so as to be away from the special locus.

	The excitations with $\omega = \hat\omega = 1$ are supersymmetric and were studied in \cite{ganchev.houppe.ea:2022,ganchev.houppe.ea:2022*1}. 
	In this paper, we focus on the simplest non-supersymmetric choice, $\omega = \hat\omega = 3$. 
	We thus choose the first-order perturbation to be:
	\begin{align}
		\delta^{(1)}\nu ~&=~ \alpha\, \xi^{n}\, \qty( 1 - \frac{2 + n}{1 + n} \xi^2 ) e^{3 i t} \,, \label{eq:linear_nu}
		\\
		\delta^{(1)}\lambda ~&=~ \beta\, \xi^{2 n} \qty(1-\xi^2) \qty(1 - 4 \frac{2+n}{1 + 2n}\xi^2 + \frac{(2 + n)(5 + 2 n)}{(1 + n)(1 + 2n)} \xi^4) \,  e^{6 i t} \,. \label{eq:linear_lambda}
	\end{align}

	\subsection{Second and third orders}
	\label{sub:naive_23}

	At order $\epsilon^2$, all the fields receive corrections.\footnote{The standard lore of AdS perturbation theory is that the scalars receive corrections at even order, while the metric and the other fields are corrected at odd orders. Here however, this rule does not stand. Indeed, the three-dimensional supergravity theory descends from a six-dimensional theory with an \adst{}$\times S^3$ vacuum \cite{mayerson.walker.ea:2020}, and the scalar $\lambda$ encodes for a deformation of the 3-sphere. This ``fake'' scalar is actually a metric mode, so exciting it at order $\epsilon^1$ breaks the usual distinction between even and odd orders in perturbation theory.} We determine them by solving the equations of motion and imposing the smoothness and boundary conditions described in section~\ref{sub:a_time_dependent_ansatz}. The expressions of the scalars $\nu$ and $\lambda$ are found to be
	{\small
	\begin{align}
	 	\delta^{(2)} \nu ~&=~  \alpha\beta \xi^{3n} (1-\xi^2) \left(n+1-(n+2) \xi ^2\right) \frac{ (1 + n) (1 + 2 n) -4 (n+1) (n+2) \xi ^2 +(n+2) (2 n+5) \xi ^4}{2 (1+n)^2 (1+2 n)} e^{3 i t} \,,
	 	\\
	 	\delta^{(2)} \lambda ~&=~ -\alpha^2 \xi^{2n} (1-\xi^2) \frac{\qty(n+1-(n+2)\xi^2)^2}{4(n+1)^2} e^{6 i t}\,.
	\end{align}\hspace{-.1em}
	}
	One can always add homogeneous solutions to these expressions, but since we assumed $\alpha \neq 0$, $\beta \neq 0$, the amplitude of these solutions can be reabsorbed into a redefinition of $\alpha$ and $\beta$. The constant of integration in $\lambda$ has been chosen so that the limit $\beta \to 0$ corresponds to the special locus.
	
	Similarly we can solve the equations of motion for the rest of the fields ; their expressions can be found in Appendix~\ref{sec:second_order_perturbation_theory}. At this order, apart from the oscillations of the scalars, all the fields are still time-independent. 

	The situation changes at order $\epsilon^3$. At this order, the time-independent ansatz of \cite{ganchev.giusto.ea:2022} breaks, and the fields acquire a genuine time-dependence. Solving the equations of motion for the two scalars $\nu$ and $\lambda$, we find
	\begin{align}
		\delta^{(3)} \nu ~&=~ \qty[ \alpha^3 P_1(\xi) + \alpha\beta^2 P_2(\xi) ] \, e^{3it} 
		+ i \,\alpha \,(v_1 \alpha^2 + v_2 \beta^2) \, \xi^{n} \qty(1-\frac{n+2}{n+1}\xi^2)  \, t\, e^{3it} \,, \label{eq:order3nu}
		\\
		\delta^{(3)} \lambda ~&=~ \qty[ \beta^3 P_3(\xi) + \alpha^2\beta P_4(\xi)  ] \, e^{6it} \nonumber\\&\quad
		+ i \,\beta \,(w_1 \alpha^2 + w_2 \beta^2) \xi^{2 n} \qty(1-\xi^2) \qty(1 - 4 \frac{2+n}{1 + 2n}\xi^2 + \frac{(2 + n)(5 + 2 n)}{(1 + n)(1 + 2n)} \xi^4) \, t\,  e^{6 i t} \,, \label{eq:order3lambda}
	\end{align}
	where $P_i$ are polynomials in $\xi$ of degree smaller than $6(n+3)$, and the $v_i, w_i$ are constants:
	\begin{equation}
	\begin{aligned}
		v_1 ~&=~ -\frac{3 \left(24 n^3+96 n^2+102 n+29\right)}{4 (n+1)^2 (2 n+1) (2 n+3) (2 n+5)} \,,
		\\
		v_2 ~&=~ -\frac{3 \left(729 n^5+5221 n^4+14397 n^3+19039 n^2+11982 n+2832\right)}{(n+1)^2 (2 n+1)^2 (2 n+3)^2 (3 n+2) (3 n+4) (3 n+5) (3 n+7)} \,,
		\\
		w_1 ~&=~ -\frac{729 n^4+4006 n^3+7659 n^2+5998 n+1608}{2 (n+1)^2 (3 n+2) (3 n+4) (3 n+5) (3 n+7)} \,,
		\\
		w_2 ~&=~ -\frac{6 \left(2304 n^5+16360 n^4+44880 n^3+59330 n^2+37671 n+9135\right)}{(n+1)^2 (2 n+1)^2 (2 n+3)^2 (4 n+3) (4 n+5) (4 n+7) (4 n+9)} \,.
	\end{aligned}
	\end{equation}

	The dependence on time occurs through the terms $t e^{3it}$ and $t e^{6it}$ in \eqref{eq:order3nu} and \eqref{eq:order3lambda}. It is important to note that these terms grow unboundedly with time. 

	Determining the validity of solutions constructed using perturbation theory is often difficult, because they present themselves as asymptotic series that do not necessarily converge. A commonly-used criterion for the validity of a solution is that the contributions at each order are smaller than the corrections of previous orders. Applying this criterion here, we find that at large times, $t \sim \cO( \epsilon^{-1} )$, the third-order corrections grow larger than the second-order corrections, and the perturbation theory breaks down.

	Unbounded terms such as these are ubiquitous in the time-dependent perturbation theory of dynamical systems, and are called \emph{secular terms}. One is often interested in questions such as the long-time stability of the solutions of such systems, but this is precisely the regime where the presence of secular terms invalidate the perturbation theory. For this reason, many techniques have been developed to \emph{resum} secular terms \cite{Drazin:1992no}. The main lesson is that the naive expansion in powers of the amplitude is not the right way to organize the perturbation theory. Instead, these resummation techniques deal with secular terms by absorbing them into shifts of frequencies.

	We will use the Poincaré-Lindstedt resummation method in section~\ref{sec:poincare_method}, but before, it is instructive to see how one can modify by hand the naive perturbation theory to resum the secular terms appearing in \eqref{eq:order3nu} and \eqref{eq:order3lambda}.

	\subsection{Dealing with the secular terms}
	\label{sub:naive_dealing_secular}

	While techniques exist to automatically resum the secular terms, it is sometimes possible, and instructive, to do it manually from the naive perturbation theory. We are interested here in the secular terms appearing in \eqref{eq:order3nu} and \eqref{eq:order3lambda}.

	First note that the secular terms at order $\epsilon^3$ can be re-expressed in terms of the first-order excitations $\delta^{(1)} \nu$ and $\delta^{(1)} \lambda$, as:
	\begin{align}
		\delta^{(3)} \nu ~&=~ \qty[ \alpha^3 P_1(\xi) + \alpha\beta^2 P_2(\xi) ] \, e^{3it} + i t \,(v_1 \alpha^2 + v_2 \beta^2) \, \delta^{(1)} \nu \,, \label{eq:sec_nu}
		\\
		\delta^{(3)} \lambda ~&=~ \qty[ \beta^3 P_3(\xi) + \alpha^2\beta P_4(\xi)  ] \, e^{6it} + i t \,(w_1 \alpha^2 + w_2 \beta^2)\, \delta^{(1)} \lambda \,.  \label{eq:sec_lambda}
	\end{align}
	We can plug these results in the full expansion of the scalars up to third-order:
	\begin{align}
		\begin{aligned}
			\nu \, &=\,  \epsilon \delta^{(1)} \nu \, + \, \epsilon^2 \delta^{(2)} \nu \, + \, \epsilon^3 \delta^{(3)} \nu \, + \, \cO(\epsilon^4)
			\\
			&=\,  \epsilon \qty[1 + i \epsilon^2 t (v_1 \alpha^2 + v_2 \beta^2)] \delta^{(1)} \nu \, + \, \epsilon^2 \delta^{(2)} \nu \, + \, \epsilon^3 \qty[ \alpha^3 P_1(\xi) + \alpha\beta^2 P_2(\xi) ] \, e^{3it} \, + \, \cO(\epsilon^4) \,,
		\end{aligned}
		\\
		\begin{aligned}
			\lambda \, &=\,  \epsilon \delta^{(1)} \lambda \, + \, \epsilon^2 \delta^{(2)} \lambda \, + \, \epsilon^3 \delta^{(3)} \lambda \, + \, \cO(\epsilon^4)
			\\
			&=\,  \epsilon \qty[1 + i \epsilon^2 t (w_1 \alpha^2 + w_2 \beta^2)] \delta^{(1)} \lambda \, + \, \epsilon^2 \delta^{(2)} \lambda \, + \,\qty[ \beta^3 P_3(\xi) + \alpha^2\beta P_4(\xi)  ] \, e^{6it}  \, + \, \cO(\epsilon^4) \,,
		\end{aligned}
	\end{align}
	where for brevity we have not expanded the terms at order $\epsilon^2$.
	The key observation is that in these expressions, the terms multiplying $\delta^{(1)} \nu$ and $\delta^{(1)} \lambda$ can be seen as the start of the expansion of two exponentials:
	\begin{align}
		\nu \, &=\,  \epsilon \, e^{i \epsilon^2 \,t \,\qty(v_1 \alpha^2 + v_2 \beta^2)} \,\delta^{(1)} \nu \, + \dots \,,
		\label{eq:resum_nu}
		\\
		\lambda \, &=\,  \epsilon \, e^{i \epsilon^2 \,t \,\qty(w_1 \alpha^2 + w_2 \beta^2)}\, \delta^{(1)} \lambda \, + \, \dots\,. 
		\label{eq:resum_lambda}
	\end{align}

	Written in this way, the expansion no longer contains terms that grow linearly with time, so the approximation remains accurate at large times. The secular terms have been \emph{resumed} into exponentials. These exponentials can be understood as shifts of the resonant frequencies of the scalars due to the non-linearities of the system. In this interpretation, the frequencies have their own asymptotic expansion: $\omega = 3 + \epsilon^2 \delta\omega_2 + \dots$, and  $\hat\omega = 3 + \epsilon^2 \delta\hat\omega_2 + \dots$. We found by resumming the secular terms that the first corrections of the frequencies are:
	\begin{equation}
	    \delta\omega_2 ~=~ v_1 \alpha^2 + v_2 \beta^2 \,,\qquad 
	    \delta\hat\omega_2 ~=~ \frac12 (w_1 \alpha^2 + w_2 \beta^2) \,.
	    \label{eq:freq_shift}
	\end{equation}

	We give a physical interpretation for these frequency shifts in section~\ref{sub:freq_shift}, but let us first mention what happens at fourth and higher orders in perturbation theory.

	\subsection{Higher orders} 
	\label{sub:higher_orders}
	

	The perturbation theory at fourth and higher orders becomes too complicated to carry out with a general integer $n$. We can however fix the value of $n$ and continue the analysis. We have done this exercise with $n=2$, up to order $\epsilon^{10}$. We have found no obstacle to continuing the expansion to even higher orders, but the increasing need for computational power. At any order, the regular solutions are always polynomials of both the radius $\xi$, and the time $t$. The degrees of these polynomials, however, increase rapidly with the order of the perturbation theory.

	We observe that the number of secular terms increases as well, as one progresses to higher orders, and resumming them also becomes more challenging. We start with order $\epsilon^4$, with $n=2$. For the scalar $\nu$, we find
	\begin{equation}
		\secterm{\delta^{(4)} \nu} ~=~ \qty(\frac{11\alpha^2\beta}{95256} - \frac{7454\beta^3}{4690310625}) i t (\delta^{(1)} \nu) \ +\ \qty(2 \delta^{(2)}\hat\omega - \delta^{(2)}\omega) i t (\delta^{(2)} \nu) \,
		\label{eq:fourth_order_nu}
	\end{equation}
	where we use $\secterm{\cdot}$ to denote the secular terms. The first term can be resummed by fixing the shift in the frequency $\omega$ at third order:
	\begin{equation}
		\delta\omega_3 ~\equiv \frac{11\alpha^2\beta}{95256} - \frac{7454\beta^3}{4690310625} \,.
	\end{equation}
	The second term, however, cannot be easily resummed. Indeed, re-expanding the resummed exponential of \eqref{eq:resum_nu} leads at fourth order to the following term:
	\begin{equation}
		\delta\omega_2 \ i t (\delta^{(2)} \nu) \,,
	\end{equation}
	which has a different prefactor than the secular term found in \eqref{eq:fourth_order_nu}. Nonetheless, the fact that the functional form of the secular term is correct, and that its prefactor can be expressed in terms of the frequency shifts, hints at some structure. The same structure can be found in the scalar $\lambda$:
	\begin{equation}
		\secterm{\delta^{(4)} \lambda} ~=~ \frac{266009\alpha^2\beta}{2894591700} i t (\delta^{(1)} \lambda) \ +\ 2 \delta\omega_2 \ i t (\delta^{(2)} \lambda) \, \,.
		\label{eq:fourth_order_lambda}
	\end{equation}
	Again, the first secular term is easily resummed by fixing the shift of $\hat\omega$:
	\begin{equation}
		\delta\hat\omega_3 ~\equiv~ \frac{266009\alpha^2\beta}{5789183400} \,,
	\end{equation}
	while the second term has the correct functional form but not the correct prefactor to be resummed.

	The order $\epsilon^5$ sees the appearance of quadratic secular terms:
	\begin{align}
		\secterm{\delta^{(5)} \nu}  \!\!&=\, - \frac12 (\delta\omega_2)^2 t^2 (\delta^{(1)} \nu) \, +\, \delta\omega_4\ i t (\delta^{(1)} \nu) \, +\, \qty(2 \delta\hat\omega_3 - \delta\omega_3) i t (\delta^{(2)} \nu)  \, +\, \delta\omega_2\ i t (\delta^{(3)} \nu) \,,
		\\
		\secterm{\delta^{(5)} \lambda}  \!\!&=\, - \frac12 (2\delta\hat\omega_2)^2 t^2 (\delta^{(1)} \lambda) \, +\, 2\delta\hat\omega_4\ i t (\delta^{(1)} \lambda) \, +\, 2 \delta\omega_3\ i t (\delta^{(2)} \lambda)  \, +\, 2\delta\hat\omega_2\ i t (\delta^{(3)} \lambda) \,,
	\end{align}
	where we have already fixed the frequency shifts:
	\begin{align}
		\delta\omega_4 \,&\equiv\, \frac{9420494051 \alpha ^4}{1287241956000}+\frac{12285630600811 \alpha ^2 \beta ^2}{16594694216100000}+\frac{3707341139902791151 \beta ^4}{245916289576471263750000} \,,
		\\
		\delta\hat\omega_4 \,&\equiv\, \frac{12920614978529 \alpha ^4}{1775745170199000}+\frac{357152922762866137 \alpha ^2 \beta ^2}{495965626036580700000}+\frac{69899075495524690733 \beta ^4}{4807663461220013206312500} \,.
	\end{align}
	Some structure appears once again in the secular terms. In each scalar, the first, second and fourth secular terms are resummed by the frequency shifts. The third secular term of both scalars, however, is not, presenting the correct functional form but not the correct prefactor.

	As a general rule, the secular terms appear to grow as $t^k$ at orders $2k+1$ and $2k+2$. Up to order $\epsilon^5$, only $\nu$ and $\lambda$ have secular terms, but they manifest in the other fields as well after order $\epsilon^6$. If they can't be resummed, the regime of validity of the approximation shrinks to shorter and shorter periods of time as one unfolds the perturbation theory to higher orders. This issue explains the need for a general method for secular term resummation, which is the subject of section~\ref{sec:poincare_method}.

	\subsection{Interpretation of the frequency shifts}
	\label{sub:freq_shift}

	We now interpret the frequency shifts \eqref{eq:freq_shift} in light of the discussion around the special locus of \cite{ganchev.giusto.ea:2023}.
	It is important to remember that the raw values computed in \eqref{eq:freq_shift} do not hold any physical meaning. As we saw in section~\ref{ssub:the_gauge_fields}, at the classical level the theory possesses a gauge invariance: for any real $\Delta\omega \in \RR$, the theory is invariant under
	\begin{equation}
	    \Phi_1 \to \Phi_1 - \frac{\Delta\omega}2 \,,\qquad \nu \to e^{i \Delta\omega \,t}\nu \,,\qquad \lambda \to  e^{2 i \Delta\omega \, t} \lambda \,,
	\end{equation}
	which has the effect of shifting both frequencies $\omega$ and $\hat\omega$ by $\Delta\omega$.
	This transformation is in fact anomalous in the quantum theory, and in the context of \cite{ganchev.giusto.ea:2022,ganchev.giusto.ea:2023}, it has been possible to compute physical frequencies, by first determining the CFT dual of the bulk excitations. While it would be interesting to know whether this procedure can also be carried out with the solutions constructed here, this is out of the scope of this paper.

	There is, however, a physical quantity that one can easily extract from the computation. While the individual frequencies shifts in \eqref{eq:freq_shift} are not gauge invariant, their difference is: 
	\begin{equation}
	    \delta\omega_2 - \delta\hat\omega_2 ~=~ \qty( v_1 - \frac{w_1}2) \alpha^2 +  \qty(v_2 - \frac{w_2}2) \beta^2 \,. \label{eq:freq_diff}
	\end{equation}
	We first notice that this quantity is generically non-zero: since it is gauge invariant, this means that there is no gauge in which these microstrata solutions are time-independent. This confirms the conjecture made in section~\ref{sub:sploc}.

	In the limit $\beta \to 0$, the solution corresponds to the special locus, and is known to be time-independent. One should recover this property through the computation presented here. This check is in fact not completely trivial, as the quantity \eqref{eq:freq_diff} does not vanish when $\beta = 0$. Looking more carefully at the solution, one should note that the secular term present at third order in $\lambda$, in equation \eqref{eq:order3lambda}, vanishes when $\beta = 0$. This means that the shift in the frequency, $\delta^{(2)} \hat\omega$, obtained by resumming this secular term, is meaningless in the limit $\beta \to 0$, thus the limit is not continuous. To obtain the right result, one needs to first fix $\beta = 0$ strictly, then restart the resummation of the non-zero secular terms. One then finds that all secular terms can be resummed, and the correct value for the shift is $\delta^{(2)} \hat\omega=\delta^{(2)} \omega$. Thus the special locus solution is indeed time-independent.

	\section{The Poincaré-Lindstedt method}
	\label{sec:poincare_method}

	We have found that the naive perturbation theory contains many secular terms, growing unboundedly with time, and restricting the regime of validity of the approximation to short times. These terms are a staple of time-dependent perturbation theory, especially when considering systems with resonances. Several resummation schemes have been developed to deal with these terms (see for example \cite{balasubramanian_holographic_2014,chen:1995ena,craps_renormalization_2014}), with the goal of constructing perturbative solutions that are valid at all times. One of the simplest  is the Poincaré-Lindstedt method, which is a way of formalizing the observations of section~\ref{sub:naive_dealing_secular}. While it is easy to implement, it sometimes fails to resum all the secular terms.\footnote{More elaborate techniques, such as the multiple scale methods, can then be used.}

	\vspace{.5em}
	Let us first illustrate the Poincaré-Lindstedt method with a simple example, the anharmonic oscillator:
	\begin{equation}
		\left\{
	    \begin{aligned}
	    	& \partial_t^2 x + x + \epsilon x^3 = 0 \,,\\
	    	& x(0) = 1 \,,\quad \partial_t x(0) = 0 \,,
	    \end{aligned}
	    \right.
	\end{equation}
	where $\epsilon$ is small. To solve this system, naively, one constructs a perturbative solution of the form $x(t) = x_0(t) + \epsilon \,x_1(t) + \dots$, and expands the equation in powers of $\epsilon$. The result up to order $\epsilon^1$ is given by
	\begin{equation}
	    x(t) = \cos(t) + \epsilon\qty[\frac1{32} \qty(\cos3t - \cos t) - \frac38 t \sin t] + \dots \,.
	\end{equation}
	The situation is now familiar: we find a secular term ``$t \sin t$'', that grows without bounds with time. At large times, when $t \propto \epsilon^{-1}$, it becomes larger than the term at zeroth-order, and renders the approximation invalid.

	It is however possible to organize the perturbation theory differently, and avoid the issue. Define a rescaled time coordinate $\tau$, and give the scaling parameter its own perturbative expansion:
	\begin{equation}
	    \tau \equiv \omega t  \,,\qq{with} \omega = 1 + \epsilon\, \omega_1 + \dots \,.
	\end{equation}
	One can now build the perturbative solution using this new time: $x(\tau) = x_0(\tau) + \epsilon \,x_1(\tau) + \dots$. In the equation, one needs to convert the time derivatives: $\partial_t \to \omega \partial_\tau$. The system becomes:
	\begin{equation}
		\left\{
	    \begin{aligned}
	    	& \omega^2 \partial_\tau^2 x(\tau) + x(\tau) + \epsilon \, x(\tau)^3 = 0 \,,\\
	    	& x(0) = 1 \,,\quad \partial_{\tau} x(0) = 0 \,,
	    \end{aligned}
	    \right.
	\end{equation}
	and the solution at first order is now given by:
	\begin{equation}
	    x(\tau) = \cos(\tau) + \epsilon\qty[\frac1{32} \qty(\cos3\tau - \cos \tau) + \qty(\omega_1 - \frac38) \tau \sin \tau] + \dots \,.
	\end{equation}
	In this expansion, the secular term is multiplied by $(\omega_1 - 3/8)$, and can be cancelled by making the choice $\omega_1 = 3/8$. With this choice, the term at order $\epsilon^1$ always remains smaller than the zeroth-order term, and the resulting perturbative solution is valid at all times. Rewriting this solution using the original time coordinate $t$, we find as expected a shift in the frequency of oscillations.

	\subsection{Normalizable excitations}
	\label{sub:nomralizable_exc_poincare}

	The simple example of the anharmonic oscillator is useful to understand the Poincaré-Lindstedt method, but is not directly applicable to our system. Indeed, the ansatz \eqref{eq:list_func_ansatz} that we work with contains two scalar fields, oscillating at two different frequencies. To adjust to this situation, we introduce not only one but two new time coordinates:
	\begin{equation}
	    T_1 = \omega \, t \,,\quad T_2 = 2\, \hat\omega\, t \,,\qq{where} \omega = \omega_0 + \epsilon^2 \delta\omega_2 + \dots \,,\quad \hat\omega =  \hat\omega_0 + \epsilon^2 \delta\hat\omega_2 + \dots \,.
	\end{equation}
	As in section~\ref{sub:linearized_equations}, we chose to study solutions whose zeroth-order frequencies are $\omega_0 = \hat\omega_0 = 3$.

	All the fields of the ansatz \eqref{eq:list_func_ansatz} now become functions of $(T_1, T_2, \xi)$, and in the equations of motion we replace the time derivatives following the rule $\partial_t \to \omega \partial_{T_1} + \hat\omega \partial_{T_2}$. We then expand the fields as
	\begin{equation}
	    X(T_1, T_2,\xi) = X^\ads + \sum_{k=1}^\infty \epsilon^k \, \delta^{(k)}\! X (T_1, T_2,\xi) \,, \quad X \in \cF \,,
	\end{equation}
	where $X^\ads$ denotes the fields of the same \ads{} background as previously, given in \eqref{eq:ads_background_fields}, and $\epsilon$ is the parameter controlling the expansion.

	The first step is to determine the normalizable excitations. Because we have two time variables, there are many more such excitations than in the previous, naive perturbation theory. Indeed, note that at order $\epsilon^1$ one has $2T_1 = T_2$, so one can multiply the excitations \eqref{eq:wave_solutions} by any function of $2T_1 - T_2$ -- the product will still solve the linearized equations of motion.

	Concretely, we can write the normalizable solutions in terms of four integers, $m, \hat m, k, \hat k \in \ZZ$. They take the form:
	\begin{align}
		\nu ~&=~ \alpha_{k,m}\, \xi^{n}\, {}_2 F_1\qty(\frac{1-3m}{2},\, n + \frac{1+3m}{2},\, 1+ n \,;\, \xi^2) \, e^{i (m T_1 + k(2T_1 - T_2))} \,,
		\\
		\lambda ~&=~ \beta_{\hat k, \hat m}\, \xi^{2 n} \, {}_2 F_1\qty(-3 \hat m	,\, 2 n + 3\hat m,\, 1+ 2n \,;\, \xi^2)\, e^{i (\hat m T_2 + \hat k(2T_1 - T_2))} \,.
	\end{align}

	The objective is to study the same family of solutions as in the naive perturbation theory of section~\ref{sub:linearized_equations}, and so we need to fix the amplitudes of the excitations at the linear order. We will thus choose to only keep the excitations with $m = \hat m = 1$, and $k = \hat k = 0$. The latter choice is somewhat arbitrary, as we expect that different choices lead to the same result after resumming and rewriting the solutions in terms of the variable $t$: the terms are simply grouped differently. We have:
	\begin{align}
		\delta\nu^{(1)} ~&=~ \alpha\, \xi^{n}\, \qty( 1 - \frac{2 + n}{1 + n} \xi^2 ) e^{i T_1} \,, \label{eq:linear_nu_poincare}
		\\
		\delta\lambda^{(1)} ~&=~ \beta\, \xi^{2 n} \qty(1-\xi^2) \qty(1 - 4 \frac{2+n}{1 + 2n}\xi^2 + \frac{(2 + n)(5 + 2 n)}{(1 + n)(1 + 2n)} \xi^4) \,  e^{i T_2} \,, \label{eq:linear_lambda_poincare}
	\end{align}
	where we have renamed  $\alpha_{0,1} \equiv \alpha$ and $\beta_{0,1} \equiv \beta$.

	\subsection{Correcting the frequency} 
	\label{sub:correcting_the_frequency}

	In what follows we fix $n=2$. At order $\epsilon^2$, the expression for the complex scalars is:
	\begin{align}
		\delta^{(2)}\nu ~&=~ \frac1{30} \, \xi^2 (3-4\xi^2) \qty[c_1 + \alpha \beta \xi^4 \qty(5-21\xi^2 + 28 \xi^4 -12 \xi^6)] e^{i(T_2 - T_1)} \,,
		\\
		\delta^{(2)}\lambda ~&=~ \frac1{63} \xi^4 (1-\xi^2) \qty[\alpha^2\qty(-2-2\xi^2+5\xi^4 )+ c_2 \qty(5 - 16\xi^2 + 12 \xi^4)] e^{2i T_1} \,,
	\end{align}
	where $c_1$ and $c_2$ are two constants of integration.
	We see the effect of the non-linearities through the fact that the scalar $\nu$ now also depends on $T_2$ and that $\lambda$ depends on $T_1$.
	The expressions of all the fields are included in appendix~\ref{sec:second_order_poincare_lindstedt}. Note that the field $\delta^{(2)}\mu_-$ also contains a constant of integration, $c_3$.

	We now turn to order $\epsilon^3$. Once again, we solve the linearized equations for all the fields of the ansatz. We are mostly interested in the solutions for the complex scalars $\nu$ and $\lambda$. The scalar $\nu$ contains the following term:
	\begin{equation}
	    \delta^{(3)} \nu ~\supset~ \frac16 \alpha \,\xi^2 (3-4 \xi^2)  \qty[\delta\omega_2 + \frac{809}{3780}\alpha^2 + \frac{20312}{2\,627\,625}\beta^2 ] \log(1-\xi^2) e^{i T_1} \,+ \dots \,.
	\end{equation}
	This term has a logarithmic singularity at infinity. We must require that the scalars are regular, and this fixes the first correction to the frequency $\omega$:
	\begin{equation}
	    \delta\omega_2 = - \frac{809}{3780}\alpha^2 - \frac{20312}{2\,627\,625}\beta^2 \,.
	\end{equation}
	Similarly, the scalar $\lambda$ also contains a log-divergence, that can be canceled by the appropriate shift in the other frequency $\hat\omega$:
	\begin{equation}
	    \delta\hat\omega_2 = - \frac{5497}{25740}\alpha^2 - \frac{9034}{1\,191\,190}\beta^2 \,.
	\end{equation}
	These results match perfectly with the \emph{ad hoc} resummation of the naive perturbation theory \eqref{eq:freq_shift}, when $n=2$.

	\subsection{A barrier at order 4}
	\label{sub:troubles_higher_orders}

	We pursue the perturbation theory to order $\epsilon^4$. The linearized equation of motion for $\nu$ at this order contains three different frequencies: $T_1$, $T_2 - T_1$ and $3T_1 - T_2$.  Interestingly, these frequencies stay located in a narrow band around $\omega_0$, no higher frequencies appear. Solving the equation, one finds that the structure of the divergent terms is as follows:
	\begin{equation}
	\begin{aligned}
	    \delta^{(4)} \nu ~=&~ \log(1-\xi^2)\bqty{ \frac16\, \alpha\, \xi^2 \qty(3-4\xi^2)  \delta\omega_3\, e^{i T_1} \ +\ (\dots) e^{i (T_2 - T_1)} \ +\ (\dots) e^{i (3T_1 - T_2)} } \\
	    & +\ \text{regular terms},
	\end{aligned}
	\end{equation}
	where for brevity we do not include the full expressions within the parentheses. To cancel these divergences, one needs to fix the order-3 shift of frequency, and impose relations between the constants of integration $c_1$, $c_2$ and $c_3$ introduced at order $\epsilon^2$:
	\begin{equation}
	\begin{gathered}
		\delta\omega_3 = 0 \,,\quad 318312 \,\beta\, c_2 - 17153136 \,\alpha \, c_1 = 1150633 \,\alpha ^2 \beta \,, \\
		134008875 \,\alpha\beta \, c_3 + 7 \qty(371875 \,\alpha^2 + 117522 \,\beta^2)\, c_1 = -81189 \,\alpha \beta^3 \,.
	\end{gathered}
	\label{eq:rel_nu_4}
	\end{equation}

	The solution for the scalar $\lambda$ also contain divergent terms, of the form:
	\begin{equation}
	\begin{aligned}
	    \delta^{(4)} \lambda ~=&~ \log(1-\xi^2)\bqty{ \frac1{10}\, \beta\, \xi^4 \qty(-5+21\xi^2-28\xi^4+12\xi^6)  \delta\hat\omega_3\, e^{i T_2} \ +\ (\dots) e^{2i T_1} \ +\ (\dots) e^{2i (T_2 - T_1)} } \\
	    & +\ \text{regular terms}.
	\end{aligned}
	\end{equation}
	These divergences can once again be cancelled by imposing the following relations:
	\begin{equation}
	\begin{gathered}
		\delta\hat\omega_3 = 0 \,, \\
		\quad c_1 = -\frac{1250565625 \alpha ^3 \beta +2185527438 \alpha  \beta ^3}{354564 \left(371875 \alpha ^2+117522 \beta ^2\right)} \,, \quad
		c_2 = \frac{4090625 \alpha ^4+353538 \alpha ^2 \beta ^2}{1487500 \alpha ^2+470088 \beta ^2} \,.
	\end{gathered}
	\label{eq:rel_lambda_4}
	\end{equation}
	
	Note, however, that the relations \eqref{eq:rel_nu_4} and \eqref{eq:rel_lambda_4} are incompatible. This means that one cannot cancel the divergences in both $\nu$ and $\lambda$. These divergences are the counterpart of the secular terms that one could not resum using the \emph{ad hoc} method of section~\ref{sec:time_dep_solutions}, though it is not possible to directly connect these two results. In particular, the particular structure of the secular terms found in section~\ref{sub:higher_orders} is not visible here.

	It appears that one cannot build solutions to the equations of motion that are regular and free of secular terms using the Poincaré-Lindstedt method. We discuss the consequences in the next section.

	\subsection{Are microstrata solutions stable?} 
	\label{sub:stability}

	Using the naive method of section~\ref{sec:naive_pert}, we have shown the existence of regular solutions to the equations of motion up to order 10 in perturbation theory. This strongly suggests that these microstrata solutions exist at all orders, at least in a limited range of time. Yet the Poincaré-Lindstedt method comes to a stop at order 4. What explains this apparent contradiction?
	
	One possibility is that some terms that have been dismissed during the computation should have been included. At all orders greater than one, one can in principle add a constant of integration for each normalizable excitation \eqref{eq:linear_nu_poincare}, and \eqref{eq:linear_lambda_poincare}. These constants are not included in the computation, because it is not possible to carry the computation with an infinite set of unfixed constants. It however appears unlikely that the absence of these terms explains the failure at order four: on physical grounds, we expect all these constants to vanish. A given mode of oscillation should not manifest in the perturbation theory at any order lower than the one where it is actually excited, i.e when it first appears on the right-hand side of the equations. In fact, such constants could have been added as well in the construction of the naive perturbation theory of section~\ref{sec:naive_pert}, yet it proved to be unnecessary.

	Rather than an artifact of the computation, the non-resummable secular terms could also be the consequence of a physical property of the solutions, namely their instability. Indeed, secular terms often indicate the presence of turbulence: a cascade of energy towards high frequencies, which leads to the concentration of energy on short length scales, and thus to gravitational collapse.
	This interplay has been revealed first in the analysis of spherically symmetric perturbations of $\ads{}_4$ \cite{bizon_weakly_2011}, and further investigations, both numerical and using perturbation theory, expanded upon this connection \cite{Dias:2011ss,Dias:2012tq,Maliborski:2013jca,Maliborski:2014rma,balasubramanian_holographic_2014,Buchel:2013uba,craps_renormalization_2014}.

	Microstrata solutions however differ significantly from the frameworks of the previously cited articles. First because in three dimensions, there exists an energy gap between global \adst{} and the first excited black hole states. Small perturbations below the threshold cannot collapse and form a black hole, as is the case in $\text{AdS}_4$. In \cite{Bizon:2013xha}, it was shown that such perturbations remain smooth forever, but still typically present a progressive loss of regularity, a phenomenon referred to as weak turbulence. While they do not collapse, these solutions are still unstable. 

	Second, these articles studied the perturbation of a single scalar in \ads{}, with multiple modes as initial data. The non-linear interactions then lead to the excitation of higher frequencies at higher orders in the perturbation theory, and it has been shown that the energy flows towards these higher frequencies, which drives the instability. The microstrata ansatz, however, comprises two scalars. By perturbing both scalars with a single (identical) resonnant mode, higher frequencies are not excited. Instead, the results of section~\ref{sec:poincare_method} indicate that frequencies are excited in a narrow band around the resonnant mode. The long-time behaviour of microstrata could thus differ significantly from the one of single-scalar preturbation theory.

	While historically Poincaré-Lindstedt has been the method of choice to resum secular terms in the context of \ads{} perturbation theory, more advanced resummation methods have also been used, such as the Two-Time framework \cite{Buchel:2013uba}, or renormalization group methods \cite{craps_renormalization_2014}. These methods have questioned the causal link between secular terms and the stability of the solutions. Whether these methods could be used to resum the secular terms of microstrata solutions is an interesting question.\footnote{Since the first preprint version of this manuscript, preliminary results suggest that the Two-Time framework is insufficient to resum the secular terms.} 


	\section{Final comments}
	\label{sec:final_comments}

	We have constructed perturbatively, through two different approaches, a family of time-dependent microstrata depending on two parameters. These solutions expand on the work of \cite{ganchev.houppe.ea:2021,ganchev.giusto.ea:2022,ganchev.giusto.ea:2023}, where the solutions were required to be time-independent, and consequently restricted to lie on a ``special locus''. We confirm that microstrata solutions exist away from the special locus, within the three-dimensional gauged supergravity truncation.

	Our first approach involved a naive expansion of the fields, and we found smooth solutions at each order in perturbation theory. The solutions are nonetheless plagued by secular terms, thereby making them unsuitable for describing the behavior of the system at large times. 
	We have presented a simple process to manually resum some secular terms, but which fails to resum all the secular terms at fourth and higher orders.  We find shifts in the physical frequencies of the scalars, due to the nonlinear interactions. The difference between these two shifts,  which is a gauge invariant quantity, has been computed. The fact that it is generically non-zero confirms the hypothesis of \cite{ganchev.giusto.ea:2023} that these solutions break the 2-to-1 phase locking at the heart of the ``Q-ball'' ansatz. 

	The second approach attempts to address secular terms in a systematic way. We adapted the Poincaré-Lindstedt method, by introducing two new time coordinates, and used it to resum the secular terms into shifts in the physical frequencies of the two scalars driving the perturbation. We computed the first deviations to the physical frequencies, and observed that they match with the manual resummation of the naive perturbation theory.

	Not all secular terms can be resummed with this approach, and we interpret this as a symptom of a possible instability of the system. In the context of the microstate geometry programme, the question of the stability of time-dependent microstrata is of great importance, as one aims at constructing stable geometries that are part of the black hole ensemble. Note, however, that an instability does not mean that the solution collapses to a black hole. In fact, perturbations below the black hole threshold cannot collapse in \adst{}. Instead, the solution may evolve towards a different, more complex microstate, that may or may not be describable within supergravity. A recent example of this phenomena was found in \cite{marolf.michel.ea:2017}: a class of previously-found instabilities has been understood as the sign of the transition from atypical microstates to more typical ones.

	It would be interesting to use the more advanced resummation frameworks to confirm or deny that all secular terms can be resummed, and more generally to determine whether the microstrata solutions are stable. One could also learn about the behavior of these solutions away from the perturbative regime using numerical methods.

	From the perspective of the dual CFT, the solutions described here are dual to multi-particle states, constructed from the same ingredients as in \cite{ganchev.giusto.ea:2023}. It would be worthwhile to study these states, to learn more about the behavior of non-BPS states at strong coupling. At the very least, one should be able to match some of the properties of the bulk solutions to the quantum numbers of the CFT states. The first step would be to determine the correct boundary conditions to impose on the supergravity solutions, and to fix the remaining gauge invariances. This would give access to the individual frequency shifts of the bulk scalars, and to anomalous dimensions of their corresponding operators in the CFT.

	\acknowledgments
	\vspace{-2mm}

	I thank Nicholas Warner for useful discussions and valuable comments on an early version of this paper.
	The work of AH was supported in part by the ERC Grant 787320 - QBH Structure, and in part by a grant from the Swiss National Science Foundation, as well as via the NCCR SwissMAP.


	\appendix

	\section{Second order naive perturbation theory} 
	\label{sec:second_order_perturbation_theory}

	The expressions of all the fields at second order in perturbation theory are
	{\small\allowdisplaybreaks
	\begin{align}
	 	\delta^{(2)} \nu ~&=~  \alpha\beta \xi^{3n} (1-\xi^2) \left(n+1-(n+2) \xi ^2\right) \frac{ (1 + n) (1 + 2 n) -4 (n+1) (n+2) \xi ^2 +(n+2) (2 n+5) \xi ^4}{2 (1+n)^2 (1+2 n)} e^{3 i t} \,,
	 	\\
	 	\delta^{(2)} \lambda ~&=~ -\alpha^2 \xi^{2n} (1-\xi^2) \frac{3 + 2n - 2(2+n) \xi^2}{4(1+n)(5+2n)} e^{6 i t}\,,
	 	\\
	 	\delta^{(2)} \mu_{+} ~&=~ -\xi^{2n} \frac{(1-\xi^2)}{4 (1+n)^2} \Bigg[\alpha^2 \qty((n + 1) - (n + 2) \xi^2)
	 	\nonumber\\&\qquad 
	 	- 2\beta^2 \xi ^{2 n} \frac{(1-\xi ^2)}{(2n+1)^2} \left((n+1) (n+2)-4 (n+1) (n+2) \xi ^2+(n+2) (2 n+5) \xi ^4\right)^2\Bigg] \,,
	 	\\
	 	\delta^{(2)} \mu_{-} ~&=~  c_1 -\alpha^2 \xi^{2n} (1-\xi^2) \frac{(n + 1) - (n + 2) \xi^2}{4 (1+n)^2} + \beta^2 \frac{\xi^{4n}}{2(n+1)(2n+1)^2} \Bigg[ \frac{(n+2)^2 (2 n+5)^2 \xi ^{12}}{n+1}
	 	\nonumber\\&\ 
	 	- \frac{8 (n+2)^2 (3 n (n+4)+11) \xi ^{10}}{n+1} + \frac{(5 n (2 n+7) (3 n (2 n+7)+34)+484) \xi ^8}{n+1}
	 	\nonumber\\&\ 
	 	- \frac{8 (n+2) (10 n (n+3) (4 n (n+3)+17)+183) \xi ^6}{(2 n+3)^2} + \frac{(5 n (2 n+5) (3 n (2 n+5)+16)+109) \xi ^4}{n+1}
	 	\nonumber\\&\ 
	 	- 8 (n+1) (3 n (n+2)+2) \xi ^2+(n+1) (2 n+1)^2 \Bigg] \,,
	 	\\
		\delta^{(2)} \Phi_{1} ~&=~ \frac{\beta ^2 \xi ^{4 n+2}}{4 (n+1)^2 (2 n+1)^2} \Bigg[ 5 (n+2)^2 \xi ^8 -2 \left(10 n^2+35 n+26\right) \xi ^6 \nonumber\\&\quad
		+\frac{2 \left(60 n^4+360 n^3+773 n^2+699 n+225\right) \xi ^4}{(2 n+3)^2} -2 \left(10 n^2+25 n+11\right) \xi ^2 +5 (n+1)^2 \Bigg] \,,
	 	\\
		\delta^{(2)} \Phi_{2} ~&=~ \frac{\alpha^2 \xi^{2n}}{8(1+n)^2} \qty[-(1+n)^2 + (3+8n+3n^2) \xi^2 - (6+10n+3n^2) \xi^4 + (2+n)^2 \xi^6] \nonumber\\&\quad
		+ \frac{\beta^2 \xi^{4n}}{4(1+n)^2(1+2n)^2} \Bigg[(1+n)^2 - 2(1+5n+2n^2)\xi^4 \\&\qquad\qquad
		+ \frac{2(27+11n+145n^2+72n^3+12n^4)}{(3+2n)^2} \xi^6 + (2+n)^2 \xi^{10} \Bigg] \,,\nonumber
		\\
		\delta^{(2)} \Psi_{1} ~&=~ \frac{\beta ^2 \xi ^2}{4 (n+1)^2 (2 n+1)^2 (2 n+3)^2} \Bigg[ - 9 \frac{1 - \xi^{4 n}}{1-\xi^2} 
		+ n \xi^{4n} \Big(2 \left(2 n^2+7 n+6\right)^2 \xi ^8 \nonumber\\&\quad
		-(2 n+3)^2 \left(8 n^2+29 n+22\right) \xi ^6 + \left(48 n^4+300 n^3+672 n^2+631 n+210\right) \xi ^4  \\&\quad
		- \left(32 n^4+188 n^3+404 n^2+371 n+114\right) \xi ^2 + \left(8 n^4+44 n^3+94 n^2+97 n+48\right) \Big) \Bigg] \,,\nonumber
		\\
		\delta^{(2)} \Psi_{2} ~&=~ -\frac{\alpha ^2 \xi ^{2 n+2}}{8 (1+n)^2} \qty[(1+n)(3+n)\qty(1-2\xi^2) + (2+n)^2 \xi^4 ] \nonumber\\&\quad
		+ \frac{\beta ^2 \xi ^2}{4 (n+1)^2 (2 n+1)^2 (2 n+3)^2} \Bigg[ - 9 \frac{1 - \xi^{4 n}}{1-\xi^2} +\xi^{4n}\Big( -\left(2 n^2+7 n+6\right)^2 \xi ^8 \nonumber\\&\qquad\quad
		+(2 n+3)^2 \left(3 n^2+12 n+8\right) \xi ^6+\left(-12 n^4-84 n^3-199 n^2-180 n-54\right) \xi ^4 \\&\qquad\quad
		+\left(4 n^4+28 n^3+69 n^2+72 n+18\right) \xi ^2-9 \Big) \Bigg] \,,\nonumber
		\\
		\delta^{(2)} \Omega_0 ~&=~ \frac{\left(1-\xi ^2\right) \xi ^{2 n}}{8 (n+1)^2 } \Bigg[ \alpha ^2 \left(-(n+2)^2 \xi ^4+2 (n (n+3)+1) \xi ^2-(n+1)^2\right) \nonumber\\&\quad
		- \beta ^2 \frac{\left(1-\xi ^2\right) \xi ^{2 n}}{(2 n+1)^2 } \Big( \left(2 n^2+3 n+1\right)^2+(n+2)^2 (2 n+5)^2 \xi ^8-4 (n+2)^2 (2 n (2 n+7)+5) \xi ^6 \nonumber\\&\qquad\quad
		+2 (n (n+3) (12 n (n+3)+25)+9) \xi ^4-4 (n+1)^2 (2 n (2 n+5)-1) \xi ^2 \Big) \Bigg] \,, 
		\\
		\delta^{(2)} \Omega_1 ~&=~ -\frac{\alpha ^2 \left(2 \left(\xi ^2-1\right) \xi ^{2 n+2}+3\right)}{4 (n+1)^2} + \frac{3 \beta ^2}{2 (n+1)^2 (2 n+1)^2 (2 n+3)} \Bigg[-3 \nonumber\\&\quad
		+ 2 (2 n+3) \left(1-\xi ^2\right)^2 \xi ^{4 n+2} \left((n+2)^2 \xi ^4-(2 n (n+3)+3) \xi ^2+(n+1)^2\right)\Bigg]\,,
		\\
		\delta^{(2)} k ~&=~ \frac{\xi ^2}{4 (n+1)^2 (2 n+1)^2 (2 n+3)} \Bigg[\alpha^2 (2 n+3) (1 +2  n)^2 \left(-2 \left(1-\xi ^2\right) \left((n+2) \xi ^2-n-1\right) \xi ^{2 n}+3\right) \nonumber\\&\quad
		+\beta ^2 \Big(18+\left(1-\xi ^2\right) \xi ^{4 n} \Big(5 (n+2)^2 (2 n+3) (2 n+5) \xi ^8 \nonumber\\&\qquad\quad
		-4 (n+2) (2 n+3) (5 n (2 n+7)+23) \xi ^6+2 (n (n+3) (60 n (n+3)+209)+171) \xi ^4 \nonumber\\&\qquad\quad
		-4 (n+1) (2 n+3) (5 n (2 n+5)+8) \xi ^2+5 (n+1)^2 (2 n+1) (2 n+3) \Big)\Big)\Bigg] \,.
	\end{align} 
	}

	The constant $c_1$ that appears at this order in the field $\mu_-$ is an integration constant (i.e a normalizable excitation of the field $\mu_-$), that is only fixed at fourth order in perturbation theory by a regularity condition. 
	
	\section{Poincaré-Lindstedt method at second order} 
	\label{sec:second_order_poincare_lindstedt}

	Using the Poincaré-Lindstedt method, the expressions of all the fields at second order in perturbation theory are
	{\small\allowdisplaybreaks
	\begin{align}
		\delta^{(2)}\nu ~&=~ \frac1{30} \, \xi^2 (3-4\xi^2) \qty[c_1 + \alpha \beta \xi^4 \qty(5-21\xi^2 + 28 \xi^4 -12 \xi^6)] e^{i(T_2 - T_1)} \,,
		\\
		\delta^{(2)}\lambda ~&=~ \frac1{63} \xi^4 (1-\xi^2) \qty[\alpha^2\qty(-2-2\xi^2+5\xi^4 )+ c_2 \qty(5 - 16\xi^2 + 12 \xi^4)] e^{2i T_1} \,,
		\\
	 	\delta^{(2)} \mu_{+} ~&=~ - \frac1{900} \xi^4 (1 - \xi^2) \Big[ 25 \alpha^2 (3-4\xi^2)^2 - 18 \beta^2 \xi^4(1-\xi^2) \qty(5 - 16 \xi^2 + 12 \xi^4)^2 \Big]  \,
	 	\\
	 	\delta^{(2)} \mu_{-} ~&=~ c_3 - \frac{\alpha^2}{36} \xi^4 (3-4\xi^2)^2(1-\xi^2) + \frac{\beta^2}{22050} \xi^8 \Big[ 63504 \xi ^{12}-294784 \xi ^{10} \nn\\&\quad +562716 \xi ^8-564768 \xi ^6+314041 \xi ^4-91728 \xi ^2+11025 \Big] \,,
	 	\\
		\delta^{(2)} \Phi_{1} ~&=~ \frac{\beta^2}{44100} \xi^{10} \qty(2205 - 9898 \xi^2 + 17110 \xi^4 -13328 \xi^6 + 3920 \xi^{8}) \,,
	 	\\
		\delta^{(2)} \Phi_{2} ~&=~ \frac{\alpha^2}{72} \xi^4 \qty(-9 + 31 \xi^2 - 38 \xi^4 + 16 \xi^6) + \frac{\beta^2}{44100} \xi^{10} \qty(441 - 1862 \xi^2 + 3194 \xi^4 -2548 \xi^6 + 784 \xi^{8}) \,,
		\\
		\delta^{(2)} \Psi_{1} ~&=~ - \frac{\beta^2}{44100} \xi^2 (1-\xi^2) \qty(9 +18 \xi^2 + 27 \xi^4 +36 \xi^6 -2160 \xi^{8} + 6816\xi^{10} - 7840\xi^{12} + 3136\xi^{14}) \,,
		\\
		\delta^{(2)} \Psi_{2} ~&=~  -\frac{\alpha^2}{72} \xi^4 \qty(15 - 30 \xi^2 +16 \xi^4) \nn\\&\quad
		- \frac{\beta^2}{44100} \xi^{2} \qty(784 \xi ^{16}-2156 \xi ^{14}+2074 \xi ^{12}-726 \xi ^{10}+9 \xi ^8+9 \xi ^6+9 \xi ^4+9 \xi ^2+9) \,,
		\\
		\delta^{(2)} \Omega_0 ~&=~ - \frac1{1800} \xi^4 (1-\xi^2) \Big[ 25 \alpha^2 (9-22\xi^2+16\xi^4) \nn\\&\qquad
		+ \beta^2 \xi^4 \qty(225 -1425 \xi^2 + 4178 \xi^4 -6054 \xi^6 +4432 \xi^{8} -1296\xi^{10} )  \Big]\,, 
		\\
		\delta^{(2)} \Omega_1 ~&=~ \frac1{36} \alpha^2 (-3 + 2\xi^6 - 2\xi^8) - \frac{\beta^2}{350} \qty(1 + \frac{14}3 \xi^{10}(1-\xi^2) (-9+32\xi^2 - 39\xi^4 + 16 \xi^6) ) \,,
		\\
		\delta^{(2)} k ~&=~  \frac{1}{36} \alpha ^2 \left(8 \xi ^8-14 \xi ^6+6 \xi ^4+3\right) \xi ^2 \nn\\&\quad
		+\frac{\beta ^2}{6300} \xi ^2 \left(-5040 \xi ^{18}+19936 \xi ^{16}-31418 \xi ^{14}+24754 \xi ^{12}-9807 \xi ^{10}+1575 \xi ^8+18\right)  \,,
	\end{align} 
	where $c_1$, $c_2$ and $c_3$ are constants of integration.
	}


	\newpage
	\bibliographystyle{jhep}
	\bibliography{microstates}

\end{document}